\documentclass[usenatbib,natbib2,natbibmnfix]{mn2e}
\usepackage{epsfig}
\usepackage{amsmath}
\usepackage{natbib}
\usepackage{times}

\newcommand{\solarmass}{$\rm h^{-1}M_\odot$}

\newcommand{\mpch}{$\rm h^{-1}Mpc$~}
\newcommand{\keV}{$\rm keV$~}
\newcommand{\phaseunit}{$\rm h^2 M_\odot~{\rm kpc}^{-3}~({\rm km~s}^{-1})^{-3}$~}
\newcommand{\gtsima}{\hbox{ $\; \buildrel > \over \sim \;$}}
\newcommand{\ltsima}{\hbox{ $\; \buildrel < \over \sim \;$}}
\def\gsim { \lower .75ex \hbox{$\sim$} \llap{\raise .27ex \hbox{$>$}}}
\def\lsim { \lower .75ex \hbox{$\sim$} \llap{\raise .27ex \hbox{$<$}}}
\title[phase space density of fermionic dark matter haloes]
{The phase space density of fermionic dark matter haloes}
\author[Shi Shao et al.]
       {Shi Shao$^{1}$\thanks{Email:shaoshi@bao.ac.cn}, 
        Liang Gao$^{1,2}$,
        Tom Theuns$^{2,3}$, and
        Carlos S. Frenk$^2$
\\
$^1$The Partner Group of Max Planck Institute for Astrophysics,
National Astronomical Observatories, Chinese Academy of Sciences, Beijing, 100012, China\\
$^2$Institute of Computational Cosmology, Department of Physics,
University of Durham, Science Laboratories, South Road, Durham DH1
3LE \\
$^3$Department of Physics, University of Antwerp, Campus
Groenenborger, Groenenborgerlaan 171, B-2020 Antwerp, Belgium\\
}

\begin{document}
\maketitle
\begin{abstract}
  We have performed a series of numerical experiments to investigate
  how the primordial thermal velocities of fermionic dark matter
  particles affect the physical and phase space density profiles of
  the dark matter haloes into which they collect. The initial particle
  velocities induce central cores in both profiles, which can be
  understood in the framework of phase space density theory.  We find
  that the maximum coarse-grained phase space density of the simulated
  haloes (computed in 6 dimensional phase space using the {\small
    EnBid} code) is very close to the theoretical fine-grained upper
  bound, while the pseudo phase space density, $Q\sim\rho/\sigma^3$,
  overestimates the maximum phase space density by up to an order of
  magnitude. The density in the inner regions of the simulated haloes
  is well described by a \lq pseudo-isothermal\rq~ profile with a
  core. We have developed a simple model based on this profile which,
  given the observed surface brightness profile of a galaxy and its
  central velocity dispersion, accurately predicts its central phase
  space density. Applying this model to the dwarf spheroidal
  satellites of the Milky Way yields values close to 0.5~keV for the
  mass of a hypothetical thermal warm dark matter particle, assuming
  the satellite haloes have cores produced by warm dark matter free
  streaming. Such a small value is in conflict with the lower limit of
  1.2~keV set by observations of the Lyman-$\alpha$ forest. Thus, if
  the Milky Way dwarf spheroidal satellites have cores, these are
  likely due to baryonic processes associated with the forming galaxy,
  perhaps of the kind proposed by Navarro, Eke and Frenk and seen in
  recent simulations of galaxy formation in the cold dark matter
  model.
\end{abstract}


\section{Introduction} \label{sect:intro}

The standard cosmological model, the ``$\Lambda$ cold dark matter
model'' ($\Lambda$CDM) has been tested over a huge range of scales,
from the entire observable universe, probed by measurements of
temperature anisotropies in the cosmic microwave background radiation
\citep[CMB;][]{Komatsu11}, to the scales of galaxy clusters and
individual bright galaxies, probed by large galaxy and Lyman-$\alpha$
forest surveys \citep{Colless05,Seljak05,Zehavi11}. On smaller
scales than this, there is no strong evidence to support the standard
model. Yet, it is on such scales that the nature of the dark matter is
most clearly manifest.  In the standard model the dark matter consists
of cold particles, such as the lightest stable particle predicted by
Supersymmetry. There are, however, models of particle physics that
predict lighter particles, such as sterile neutrinos, that would
behave as warm (WDM), rather than cold dark matter (CDM) \citep[see]
[for discussions of recent experimental
constraints]{Feng10,Hooper12}. No current astronomical data can
distinguish between these alternatives.

If the dark matter particles freeze out while in thermal equilibrium,
their kinematics leave a predictable imprint in the power spectrum of
primordial perturbations \citep{Zeldovich65}. Hot dark matter
particles, such as light neutrinos, decouple while still
relativistic; their large thermal velocities dampen fluctuations below
a \lq free streaming\rq\ length, $\lambda_{\rm FS}$, which is of the
order of a few tens of megaparsecs at redshift $z=0$
\citep{Bond_Szalay83}.  Early N-body simulations showed that this is
too large to be compatible with the level of clustering measured in
the galaxy distribution \citep{Frenk83}, thus ruling out light
neutrinos as the dominant form of dark matter. Recent analyses of the
power spectrum of the galaxy distribution \citep{Cole05}, the
Lyman-$\alpha$ forest \citep{Viel10} and the CMB \citep{Komatsu11}
constrain the sum of the neutrino masses to be $\sum m_\nu < 0.58$~eV.

Cold dark matter particles decouple after they have become
non-relativistic. For a typical cold, weakly interacting, massive
particle, $\lambda_{\rm FS}$ is of the order of a parsec and the
corresponding \lq Jeans\rq\ mass is of the order of $10^{-6}M_\odot$
\citep{Green05}; free streaming in this case is not relevant for
galaxy formation. The intermediate case of warm dark matter
corresponds to particles that decouple while still relativistic, yet
become non-relativistic before the epoch of radiation-matter equality.
Their free streaming scale could then be of order the size of a
galaxy, in which case they can affect the build-up of dark matter
haloes and of the galaxies forming within them
\citep{Bode01,Lovell12}, as well as the formation of the first stars
\citep{Gao07}. Dark matter particles need not freeze out in thermal
equilibrium, as is the case, for example, of the axion and the sterile
neutrino discussed by \cite{Boyarsky09}. In this case, the relation
between the dark matter particle mass and $\lambda_{\rm FS}$ is more
complex. \cite{Boyarsky09b} discuss current limits on such WDM models.

Except perhaps in the central regions of galaxies, the dynamics of
dark matter particles are driven purely by their own gravity and are
thus governed by the Vlasov-Poisson equations. Then, according to
Liouville's theorem, the distribution function, $f({\bf x},{\bf
  v},t)$, of the particles -- the mass per unit volume in phase space
-- is time-independent, $Df({\bf x},{\bf v},t)/Dt=0$. Phase space
mixing, or coarse-graining, can only decrease the phase space density
below this fine-grain bound set by the nature of the particles.
\cite{Tremaine79} used this property to constrain the nature of
stable leptons as the dominant form of dark matter. Since the
intrinsic velocities of CDM particles are small, their fine-grained
phase space density limit is very high and such haloes are cuspy
\citep[e.g.][]{Navarro96b, Navarro97, Diemand08,Stadel09, Navarro10,
  Gao12}. For WDM (and {\em a fortiori} for HDM), the phase space
density bound is much lower and such haloes develop central cores
\citep{Hogan00,Bode01}.

Evidence for central cores in galaxies is controversial
\citep[see][and references therein for a recent
discussion]{Frenk_White12}. 
Recently, \cite{Walker_Penarrubia12} have argued that
the kinematics of the Fornax and Sculptor dwarf spheroidals in the
Milky Way rule out central density cusps, but
\cite{Strigari_Frenk_White10} have shown that the data for these and
other dwarfs are consistent with the cuspy profiles seen in cold dark
matter simulations. On the other hand, \cite{Boylan11}, \cite{Lovell12} and
\cite{Parry12} have shown that the central concentration of the dark
matter haloes of nine dwarf spheroidals are lower than expected in 
simulations of cold dark matter haloes \citep{Springel08}
(but see \cite{Wang12}). Although baryonic processes associated with
the forming galaxy could lower the central density of haloes and even
induce a core \citep{Navarro96a,Read05,Mashchenko08,Pontzen12}, it is
also possible, in principle, that the kind of phase space constraints
just discussed could be at work at the centres of the dwarf
spheroidals.

In this paper we investigate the effect of primordial dark matter
particle thermal velocities on halo density profiles. To this aim, we
carry out N-body simulations from initial conditions in which the
power spectrum has a small-scale cut-off and the particles representing
the dark matter have significant thermal velocities. Our setup does not
correspond exactly to any particular HDM or WDM particle
candidate. Rather, we are interested in the more general problem of
how phase space constraints are satisfied in cosmological N-body
simulations. In particular, we determine the level at which the
\cite{Tremaine79} bound is satisfied. We then model our numerical
results, generalise them to the case of specific WDM 
candidates, and apply them to Milky Way satellites.

As we were completing this work, \citet{Maccio12} published a paper
investigating similar issues to those of interest here, using similar
numerical techniques. They interpret their results in terms of a
macroscopic pseudo phase space density (defined in Eqn~10 below),
which, as we show, overestimates the coarse-grained phase space
density by up to an order of magnitude. \footnote{As noted in an {\em
    Erratum} \citep{Maccio12E}, the velocity dispersion of the dark
  matter particles in the initial conditions of their simulations is
  incorrect by a factor of $\sqrt3$.}
The remainder of this paper is structured as follows. In Section~2, we
describe our set of numerical simulations. In Section~3, we briefly
review relevant aspects of phase space theory and describe our methods
for calculating the coarse-grained phase space density in the
simulations. In Section~4, we investigate phase and real space density
profiles of dark matter haloes and compare these with our theoretical
estimates. In Section~5, we use a model based on our numerical
simulations to revise the lower mass limit of WDM 
particles. Our paper concludes in Section~6 with a summary and
discussion.

\section{phase space density}

In this section, we briefly review those aspects of phase space
density theory that are relevant to general fermionic dark matter
particles.

\subsection{Fine-grained phase space density}

The evolution of a system of collisionless particles is described by
the Vlasov equation.  According to Liouville's theorem, the
fine-grained phase space density, $f(\mathbf{x},\mathbf{v},t)$ -- the
mass density in an infinitesimal six-dimensional phase space volume,
$d^3\mathbf{x}d^3\mathbf{v}$, centred on the point
$(\mathbf{x},\mathbf{v})$ at time $t$ -- is conserved: $Df/Dt=0$
(see e.g. \cite{Peebles70} for the General Relativistic description).

Following \cite{Tremaine79}, \cite{Madsen91} and\cite{Hogan00}, using
the Fermi-Dirac occupation distribution we can derive an upper limit
on ${\cal F}_{\rm FD}$, the fine-grained phase space density of a
relativistic fermionic relic in kinetic equilibrium:
\begin{eqnarray}
\label{eq:FD}
{\cal F}_{\rm FD}({\bf p}) &=& {g\over (2\pi\hbar)^3}\,\frac{1}{e^{E/k_{\rm B}T}+1}\nonumber\\
                                         &\le & {g\over 2 (2\pi\hbar)^3}\label{eq:R}\,.
\end{eqnarray}
Here, $E=[(mc^2)^2+(pc)^2]^{1/2}$ is the energy of a particle with
mass $m$ and momentum ${\bf p}$, and $g$ is the number of degrees of freedom
\citep{KolbTurner90}. This expression may be more familiar from the
derivation of the equation of state of degenerate fermions. Warm dark
matter particles are relativistic when they decouple, {\em i.e.} when
$T=T_D$, the decoupling temperature. After decoupling both the energy
and the temperature decline in proportion to $(1+z)$ and hence the
shape of the Fermi-Dirac distribution remains unchanged.  Finally,
under a Lorentz transformation, lengths contract in proportion to the
Lorentz factor, $\gamma$, whereas momenta increase in proportion to
$1/\gamma$ so the phase space density is an invariant.

The function ${\cal F}_{\rm FD}$ is the number density of particles
per unit volume in momentum space, an appropriate choice when the
particles are relativistic. When the particles become
non-relativistic, we can write the phase space density, $f_{\rm FD}$, in
terms of the {\em mass} density of particles per unit volume in {\em
  velocity} space, which introduces a factor $m^4$, so the
non-relativistic version of Eqn.~({\ref{eq:R}) becomes
\begin{eqnarray}
 f_{\rm FD}^{\rm max} &=& \frac{g\,m^4}{2(2 \pi \hbar)^3}
  \nonumber\\ 
  &=&0.42\,M_\odot\,{\rm kpc}^{-3}~({\rm km}~{\rm s}^{-1})^{-3}\,
  {g\over 2}\,({m\,c^2\over 0.03~{\rm keV}})^4\,.
  \label{eq:FD,max}
\end{eqnarray}
In the linear regime, the density of these non-relativistic particles
decays as $(1+z)^3$ but their peculiar velocities decay as $(1+z)$;
hence this bound does not evolve with redshift, as expected.

\subsection{Coarse-grained phase space density}
The coarse-grained phase space density, $F(\mathbf{x},\mathbf{v},t)$,
is defined as the mass density in a {\em finite} six-dimensional phase
space volume, $\Delta^3\mathbf{x}\Delta^3\mathbf{v}$, centered on the
point $(\mathbf{x},\mathbf{v})$ at time $t$. The averaging process
decreases the phase space density and hence $F\leq f$ (see, for
example, \cite{Tremaine86} and \cite{Mathur88} for the application of
Liouville's theorem in an astronomical setting). Phase mixing is
expected to be small in the region with the highest phase space
density, and the maximum value of $F$ for a system is therefore
expected to be close to the value of $f$ \citep{Lynden-Bell67}. Of
course, in any real system, baryonic effects may increase or decrease
the phase space density. In a simulated dark matter halo we can
estimate $F$ in 6 dimensions and verify whether indeed $F^{\rm
  max}\leq f_{\rm FD}^{\rm max}$; we will do so below.

We begin by discussing various ways in which $F$ can be estimated for
a dark matter halo.  We assume the central part of the halo to have a
pseudo-isothermal density profile \citep[e.g.][]{Kent86, Begeman91},
which is a good fit to the simulated dark matter halo discussed
below. This profile is given by

\begin{equation}
  \label{eq:density}
  \rho(r) = \frac{\rho_{0}}{1 + \left(r/r_c\right)^2},
\end{equation}
where $\rho_0$ is the core  density and $r_c$ the core radius. The
corresponding asymptotically flat circular velocity is $V_c = (4 \pi G
\rho_0 r_c^2)^{1/2}$. Assuming isotropic orbits $V_c$ is related to the
one-dimensional velocity dispersion $\sigma$, by $V_c = \sqrt2
\sigma$. Under these assumptions, central density, core radius and
velocity dispersion are related by
\begin{equation}
  \label{eq:pseudo-iso}
  \rho_{0} = \frac{1}{2 \pi G}\frac{\sigma^2}{{r_c}^2}\,.
\end{equation}

To obtain the corresponding coarse-grained phase space density, we
need to know the velocity distribution function.  Dynamically relaxed systems
often have a Maxwellian velocity distribution, and we will assume this
to be a good description for the central regions of the dark matter
halo \citep{Vogelsberger09}. The maximum density in velocity space is then
$(2\pi\sigma^2)^{-3/2}$, and therefore the maximum coarse-grained
phase space density, which occurs at the centre of the halo, is
\begin{eqnarray}
F^{\rm max}_{\rm iso} &=& \frac{\rho_0}{(2 \pi \sigma^2)^{3/2}}\nonumber\\
&=&\frac{1}{(2\pi)^{5/2} G} \frac{1}{\sigma r_c^2}\label{eq:iso,max}\\
&=& 0.06\, {\rm M_\odot}\,{\rm kpc}^{-3}~({\rm km}~{\rm s}^{-1})^{-3}\nonumber\\
&\times &({\sigma\over 100~{\rm km}~{\rm s}^{-1}})^{-1}\,({r_c\over 20~{\rm kpc}})^{-2}\,.
\end{eqnarray}
If the core is due to free-streaming of a WDM particle, then requiring
$F^{\rm max}_{\rm iso}\leq f^{\rm max}_{\rm FD}$ constrains the WDM
particle mass to be
\begin{equation}
m\,c^2\, ({g\over 2})^{1/4}\geq 8.2~{\rm keV}\,({\sigma\over {\rm
    km}~{\rm s}^{-1}})^{-1/4}\,({r_c\over {\rm pc}})^{-1/2}\,.
\label{eq:Fbound}
\end{equation}

To derive the limit in Eqn.~(\ref{eq:iso,max}) we had to make an
assumption about the form of the density profile of the central
halo. \citet{Boyarsky09} introduced a \lq coarsest\rq\ phase space
density, $F^{\rm max}_{\rm Boy}$, which corresponds to a maximal
coarse graining and does not require this assumption. Instead it makes
use of the value of the enclosed mass, $M(R)$, of the halo over the
whole available phase space volume $\Delta {\bf x}\,\Delta{\bf
  v}=(4\pi/3)^2 R^3 v_\infty^3$, where $v_\infty$ is the local escape
speed. For isotropic orbits, $v_\infty \geq \sqrt{6} \sigma$, leading
to the maximum coarse-grained value of $F$ of

\begin{equation}
  \label{eq:Boyarsky}
  F_{\rm Boy}^{\rm max} = \frac{\rho_0}{8 \pi \sqrt{6} \sigma^3} \approx \frac{3\ln 2}
  {16 \sqrt{6}\pi^{2} G} \frac{1}{\sigma r_h^2} \approx 0.18 F^{\rm max}_{\rm iso}\,,
\end{equation}
where $r_h$ is the half-light radius, defined by \citet{Pryor90},
\begin{equation}
  \label{eq:pryor}
  \rho_{0} = \frac{3 \ln2}{2 \pi G}\frac{\sigma^2}{{r_h}^2}.
\end{equation}
This calculation is less restrictive and hence $F^{\rm max}_{\rm Boy}<
F^{\rm max}_{\rm iso}$, as should be.

Finally, an often used {\em approximation} to $F$ is
\begin{equation}
  Q \equiv {\rho\over \langle v^2 \rangle^{3/2}}\,,
\end{equation} 
introduced by \citet{Hogan00} and more recently used, for example, by
\cite{Taylor01, Ascasibar04, Dehnen05, Peirani06, Hoffman07,
  Colin08, Vass09, Navarro10}. We will refer to this as a {\em pseudo}
phase space density, since it only has the dimensions of $F$, but it
is not a proper coarse-grained average of $f$. Indeed, the maximum
value of $Q$ for the halo profile of Eqn.~(\ref{eq:density}) under
the assumptions of isotropic Maxwellian velocities, is
\begin{equation}
  \label{eq:Hogan}
  Q^{\rm max} = \frac{(2 \pi)^{3/2}}{3\sqrt{3}} F^{\rm max}_{\rm iso}\approx 3.03\,F^{\rm max}_{\rm iso}\,.
\end{equation}
The upper bound to the fine-grained distribution for thermal fermions
discussed by \citet{Hogan00} is very similar to the value in
Eqn.(\ref{eq:FD,max}) above, $f^{\rm max}_{\rm Hogan}\approx 0.97
f^{\rm max}_{\rm FD}$ (see also \cite{Boyarsky09}). Requiring $Q^{\rm
  max}\leq f^{\rm max}_{\rm Hogan}$ then overestimates the constraint
on the dark matter particle mass by a factor of $(3.12)^{1/4}$.

\begin{figure*}
\centerline{\includegraphics[width=1\textwidth]{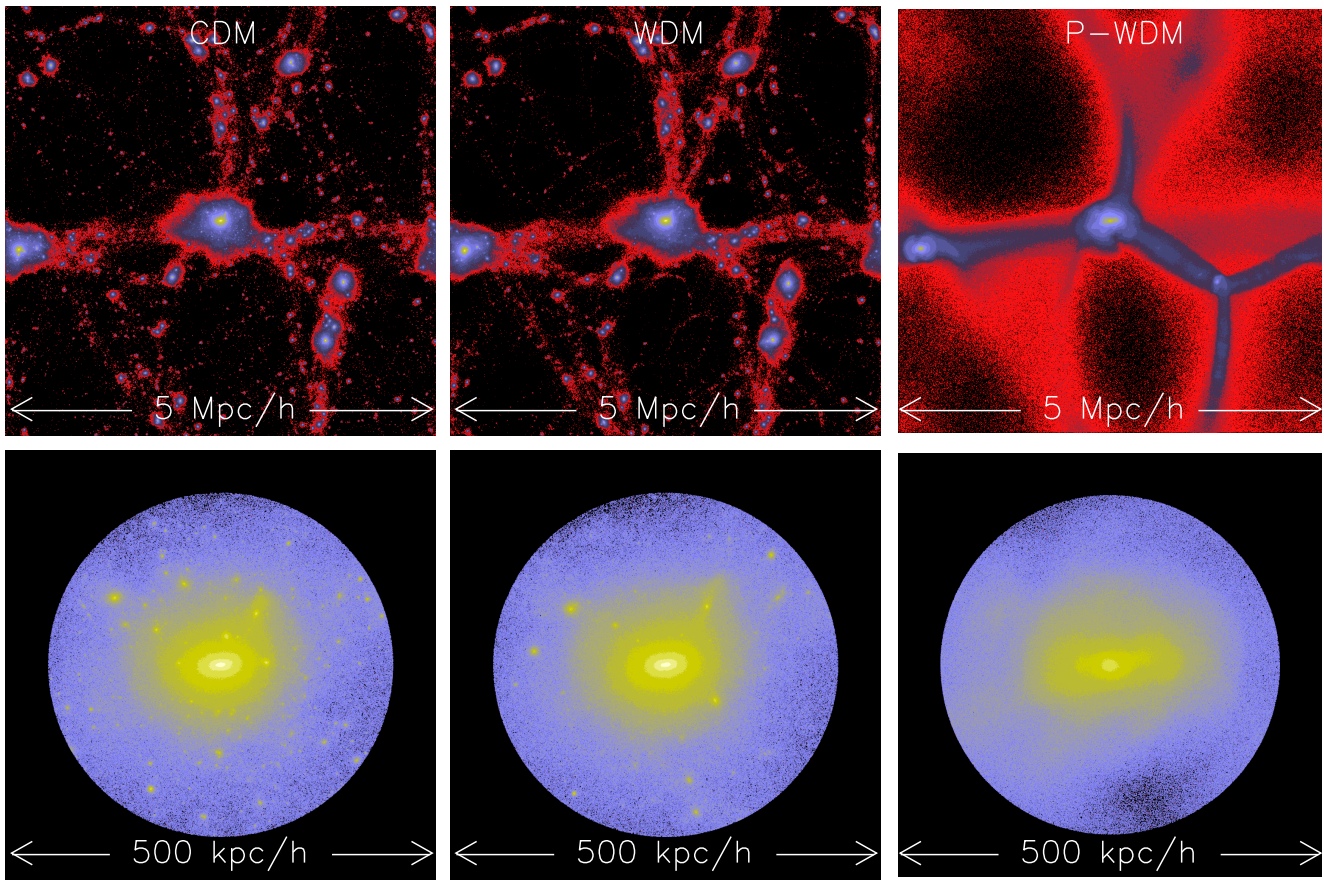}} 
\caption{
\label{fig:slice}
Projected density in a 5$h^{-1}$Mpc cube at $z=0$ for the CDM, WDM,
and P-WDM simulations (left to right). The top panels show the full
simulation volume in a slice of 1.5\mpch; the bottom panels zoom in on
the most massive halo. As expected, features on scales larger than the
free-streaming length are very similar, but on smaller scales the WDM
runs have much less substructure. Model P-WMD, which had substantial
initial random velocities, has almost no substructure at all.}
\end{figure*}

\begin{figure*}
\begin{center}
\centerline{\includegraphics[width=1.0\linewidth]{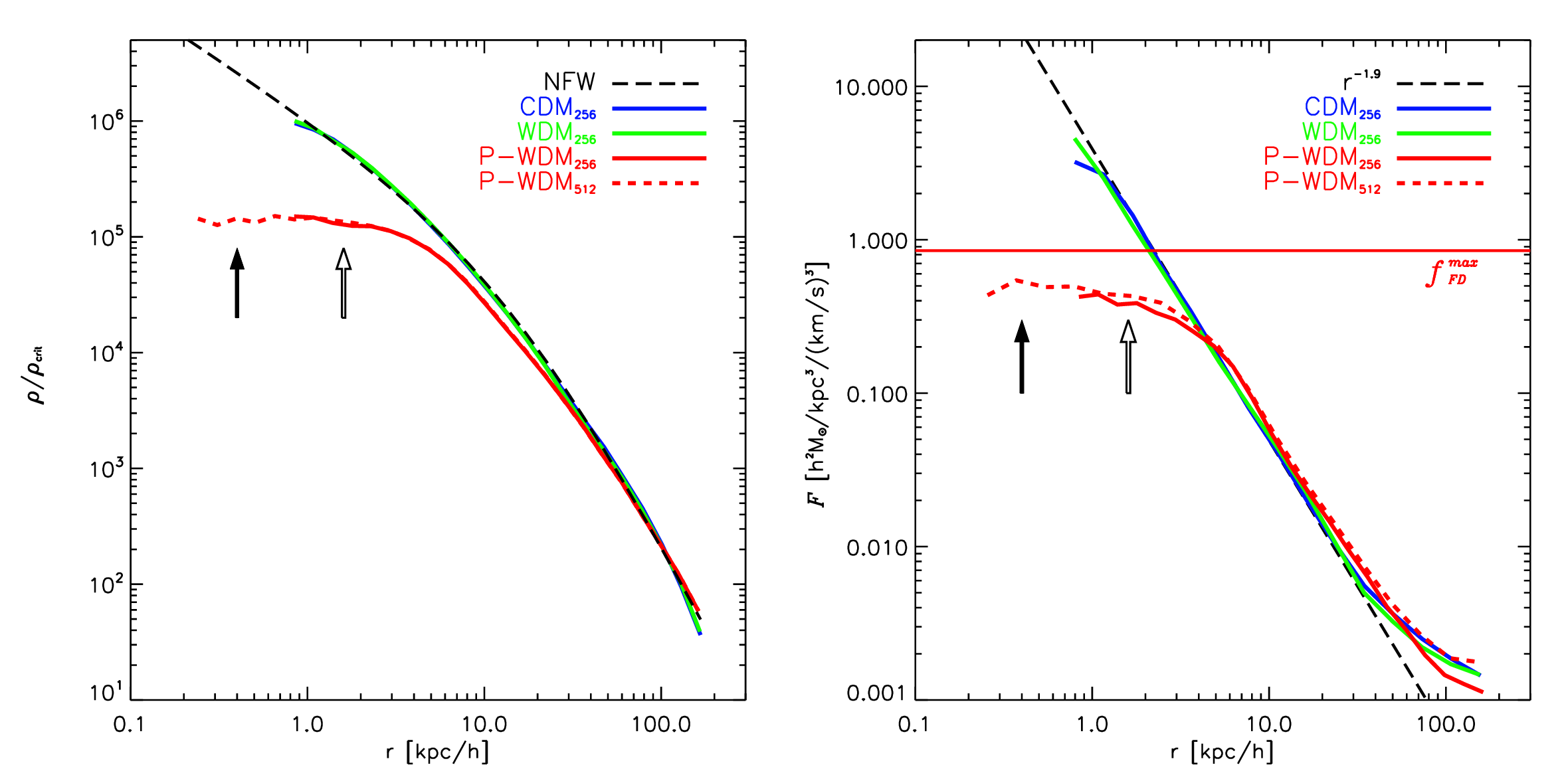}}
\end{center}
\caption{
  \label{fig:all} Mean, radially averaged density profile, $\rho(r)$
  (left), and coarse-grained phase space density profile, $F_{\rm
    NB}(r)$ (right), for the most massive halo in the CDM (blue), WDM
  (green) and P-WDM (red) simulations.  Phase space densities were
  computed using the {\small EnBiD} code. The black-dashed lines are
  NFW and power-law fits to $\rho$ and $F_{\rm NB}$ respectively for
  the CDM model.  Different line styles correspond to simulations of
  different numerical resolution, as given in the legend, with arrows
  indicating twice the corresponding softening lengths. The CDM and
  WDM profiles are nearly indistinguishable, with $\rho(r)$ and
  $F_{\rm NB}(r)$ well fit by the NFW and power-law profiles
  respectively. In contrast, the P-WDM model, which has significant
  initial thermal velocities, develops a core. The maximum value of
  the phase space density, $F_{\rm NB}$, in the P-WDM model is close
  to, but smaller than the fine-grained upper limit, $f_{\rm FD}^{\rm
    iso}$, indicated by the horizontal red line.  }

\end{figure*}

\subsection{Coarse-grained phase space density in simulations}

A robust measurement of $F$ in an $N$-body system requires a full
six-dimensional calculation. In recent years, a number of independent
approaches have been developed that allow such a calculation
\citep{Arad04, Ascasibar05, Sharma06, Vogelsberger08}. The method of
\cite{Arad04} uses the Delaunay tessellation of the $N$ particles
in 6 dimensional $({\bf x},{\bf v})$ phase space.  Two other
algorithms, {\small FiEstAS} \citep{Ascasibar05} and {\small EnBiD}
\citep{Sharma06}, are based upon the idea of a binary $k-d$-tree, {\em
  i.e.} repeated subdivisions of phase space in each of its 6
dimensions into nodes that contain (approximately) the same number of
particles until each node contains a single particle. Both the
Delaunay tessellation and the $k-d$ tree then associate a \lq
volume,\rq\ $V_i$, to a particle, and the corresponding phase space
density is simply $m_{\rm particle}/V_i$, where $m_{\rm particle}$ is
the mass of the particle in the simulation. These algorithms contain
adjustable parameters that can be chosen to improve the estimates. For
example, \cite{Ascasibar05} include a boundary correction to improve
the behaviour of the method at low density and smoothing kernels to
reduce particle noise.

In this paper we make use of the publicly available {\small EnBiD}
code \citep{Sharma06} to estimate the coarse-grained phase space
density in our simulations. {\small EnBiD} employs a {\it Shannon
  Entropy} formulation which gives accurate results in high density
regions. We have included the \lq boundary correction\rq\ but did not
use the smoothing option because it may underestimate the highest
phase space density in a dark matter halo, which is crucial in this
study. We will denote the coarse-grained phase space density of the
simulated haloes calculated using {\small EnBiD} as $F_{\rm NB}$

\section{$N$-body simulations}

\subsection{Physical requirements}

If dark matter particles have large intrinsic velocities at early
times, two distinct physical effects become important: free streaming
out of density perturbations and a maximum achievable phase space
density. In the linear regime, the WDM power spectrum of fluctuations, $P_{\rm WDM}(k)$,
is suppressed relative to the CDM power spectrum, $P_{\rm CDM}(k)$, by
a factor 
\begin{equation}
T^2(k) \equiv {P_{\rm WDM}(k)\over P_{\rm CDM}(k)}\,,
\end{equation}
where $k$ is the wavevector. Free streaming introduces a smallest
wavevector, $k_{\rm H}$, for which $T<1$, and some authors refer to
$2\pi/k_{\rm H}$ as the \lq free-streaming length\rq. However, when
considering structure formation, it is more useful to characterise the
effects of free streaming by the value of $k_{\rm
  FS}=2\pi/\lambda_{\rm FS}$ for which the amplitude of the power
spectrum is reduced by a factor two relative to the CDM case, such
that $T^2(k=k_{\rm FS})=1/2$. The shape of the function $T(k)$ depends
on the nature of the dark matter. A commonly used approximation for
thermally produced WDM particles is
\begin{equation}
T(k) = (1+(\alpha\,k)^2)^{-5}\,,
\label{eq:Tk}
\end{equation}
\citep{Bode01}. For this case,
\begin{equation}
\lambda_{\rm FS} = 23.45\,\alpha\,, \hfill \nonumber 
\end{equation}
\begin{equation}
\alpha = 0.05({\Omega_{\rm WDM}\over 0.3})^{0.15}({h\over 0.72})^{1.3}({mc^2\over {\rm keV}})^{-1.15}\,h^{-1}{\rm Mpc}\,,
\label{eq:fs}
\end{equation}
where $\Omega_{\rm WDM}$ is the total WDM density today in units of
the critical density, $m$ is the WDM particle mass, and
$H=100\,h\,{\rm km}~{\rm s}^{-1}~{\rm Mpc}^{-1}$ is the Hubble
constant at $z=0$ \citep{Bode01}. The numerical coefficient depends on
the effective number of degrees of freedom, $g_X$, of the particle and
we have assumed $g_X=1.5$. With this definition we find
\begin{equation}
\lambda_{\rm FS} = 1.2\,({\Omega_{\rm WDM}\over 0.3})^{0.15}({h\over 0.72})^{1.3}({mc^2\over {\rm keV}})^{-1.15} h^{-1}{\rm Mpc}\,;
\end{equation}
the corresponding mass is
\begin{eqnarray}
M_{\rm FS} &\equiv&{4\pi\over 3}\,\rho_m\,\lambda_{\rm FS}^3\nonumber\\
                    &=& 6\times 10^{11}\,({\Omega_{\rm WDM}\over 0.3})^{1.45}({h\over 0.72})^{3.9}({mc^2\over {\rm keV}})^{-3.45}\,h^{-1}M_\odot.\nonumber\\
                    \label{eq:Mfs}
\end{eqnarray}
We will refer to $\lambda_{\rm FS}$ as the free-streaming
scale\footnote{Unfortunately, not all authors agree on a single
  definition of $\lambda_{\rm FS}$.}

The second important effect is the limit on the phase space density
achievable by dark matter particles with significant primordial
thermal velocities. These have a Fermi-Dirac distribution,
$\left[\exp(v/v_0)+1\right]^{-1}$, with $v_0$ given by \citep{Bode01} 
\begin{equation}
\label{eq:vel}
v_0(z) \approx .012(1+z)(\frac{\Omega_{\rm WDM}}{0.3})^{1/3}(\frac{h}{0.65})^{2/3}(\frac{1.5}
{g_X})^{1/3}(\frac{\rm keV}{m_x})^{4/3} {\rm km/s}.
\end{equation}} 
These initial velocities induce a core radius, $r_c$, in collapsed
haloes which, for thermal dark matter relics, scales with the velocity
dispersion of the halo, $\sigma$, as $r_c\propto \sigma^{-1/2}
m^{-2}$ \citep{Hogan00, Bode01}.  Free streaming prevents haloes
of virial radius\footnote{We define the virial radius, $R_{200}$, of
  a cosmological dark matter halo as the radius within which the mean
  density is 200 times the critical density. The corresponding enclosed
  mass is $M_{200}$.} $R_{200}\ltsima \lambda_{\rm FS}$ from forming
in significant numbers. As we will show below, the core radius of such
haloes is $r_c\ll \lambda_{\rm FS}$.  Since $r_c\propto \sigma^{-1/2}$,
the ratio $r_c/R_{200}$ is even smaller for more massive haloes.

The small value of the ratio $r_c/R_{200}$ makes it challenging to investigate core radii
in simulations since the calculation needs to resolve the huge dynamic
range between $r_c$ and $\lambda_{\rm FS}$. This is not
currently practical for values of $\lambda_{\rm FS}$ typical of those
expected in WDM models.  However, if one is interested in the more
general problem of the relationship between $r_c$ and the phase space
density, it is not necessary for the cut-off in the power spectrum
to correspond to the thermal velocities of a  particular dark
matter candidate. This is the approach we take in this study: we
perform a numerical experiment by choosing a value of $\lambda_{\rm FS}$ but giving 
the particles much larger intrinsic velocities than would be appropriate for a WDM
particle of mass $m$ according to Eqn.~(\ref{eq:fs}). For brevity, we will
refer to these models as pseudo-WDM models (P-WDM). We stress that this is
not a self-consistent dark matter model: the resulting core radii will
be much larger than expected for a WDM particle of that mass, but this
is immaterial for our purposes. This procedure is similar to that
adopted by \citet{Maccio12}. For comparison, we also investigate a
self-consistent WDM model in which the initial thermal velocities are
compatible with the cut-off in the power spectrum. Since the thermal
velocities are now much smaller than the initial linear gravitational velocities, 
we expect their effect to be negligible.

\subsection{Simulation details}

All our simulations are of periodic cubic volumes of linear size
5$h^{-1}$~Mpc. The assumed cosmological parameters are
$(\Omega_m,\Omega_\Lambda,h,\sigma_8)=(0.3,0.7,0.7,0.9)$, but the exact
choice of these numbers is not important here. We have run two
reference simulations: a standard $\Lambda$CMD (labelled CDM) and a
self-consistent $\Lambda$WDM (labelled WDM) model.  These have $256^3$
particles, corresponding to an $N$-body particle mass of $6.2 \times
10^{5}$\solarmass. The CDM transfer function was computed using
CMBfast \citep{Seljak96}. For the WDM simulation, the power spectrum
of  the initial conditions was obtained by multiplying the CDM
spectrum by $T(k)^2$, as given by Eqn.~(\ref{eq:Tk}). The assumed
free streaming length scale, $\lambda_{\rm FS} \approx 0.5$~\mpch, corresponds
to a 2~keV thermal relic WDM particle and each particle is given the  
appropriate random velocity drawn from the Fermi-Dirac distribution 
with $v_0$ given by Eqn.~(\ref{eq:vel}). For definiteness we
will always assume the WDM particle has $g=2$ spin degrees of freedom.

The pseudo-WDM simulations (labelled P-WDM) have a similar set up to
the WDM case, except that the random velocities are much larger and
would correspond to those appropriate to a WDM particle of mass
0.03~keV. For ease of comparison, the Gaussian random field used to
initialise the simulations uses the same phases for all three models.
To enable statistical comparisons, we carried out a further 3 P-WDM
simulations using the same cosmological parameters but different
realisations of the initial conditions; these are denoted
P-WDM$_{256}$.  Finally, to investigate numerical convergence we
carried out one simulation with 8 times better mass resolution, also
with several different realisations of the initial conditions.  These
are denoted by P-WDM$_{512}$. The starting redshift for the CDM and
WDM run was $z=100$, but for the P-WDM runs it was $z=20$ since in
this case structure formation is significantly delayed by the
  larger thermal velocities. Table~\ref{tab:sims} summarises the
parameters of the simulations.  All our simulations were performed
with the {\sc Gadget-3} code, an improved version of {\sc Gadget-2}
\citep{Springel05}.

\begin{table}
  \caption{Simulation details. In all cases the assumed cosmological parameters are
    $(\Omega_m,\Omega_\Lambda,h,\sigma_8)=(0.3,0.7,0.7,0.9)$ and the
    simulation cube is $5h^{-1}$~Mpc on a side. The WDM and pseudo-WDM (P-WDM) models
    use the CDM power spectrum suppressed by $T^2(k)$ from
    Eqn.~(\ref{eq:Tk}), using the parameters for a thermal WDM particle
    of mass $mc^2=$~2~keV with $g=2$ spin degrees of freedom. The
    corresponding free-streaming length is 
    $\lambda_{\rm FS}\approx 0.5\,h^{-1}$~Mpc. The added velocities for the WDM model are
    consistent with the WDM particle properties, whereas for the P-WDM 
    run, they correspond to those of an $mc^2=0.03$~keV
    thermal WDM particle.}
\centering
\begin{tabular}{ l l  l l l}
\hline
Name & Number of  & particle mass & softening $\epsilon$ & $mc^2$\\
&     particles                             & $h^{-1}{\rm M_\odot}$      & $h^{-1}$kpc  & keV\\
\hline
CDM & $256^3$ & $6.2 \times 10^5$   & 0.8 & -- \\
WDM & $256^3$ & $6.2 \times 10^5$  & 0.8 & 2 \\
P-WDM$_{256}$ & $256^3$ & $6.2 \times 10^5$  & 0.8 & 0.03\\
P-WDM$_{512}$ & $512^3$ & $7.8 \times 10^4$ & 0.2 & 0.03\\
\hline
\end{tabular}
\label{tab:sims}
\end{table}

\subsection{Comparisons of CDM and WDM simulations}

Images of the CDM, WDM and P-WDM$_{256}$ simulations are displayed in
Fig.~\ref{fig:slice} for the fiducial $256^3$ particle run.  As
anticipated, the overall appearance of the main halo in the three
simulations is very similar, but the amount of small-scale
substructure is very different.  The cut-off in the initial power
spectrum in the WDM simulations has the effect of suppressing the
formation of structure on scales below $\lambda_{\rm FS}$. This is the
reason why there are far fewer substructures in the WDM than in the
CDM images.  The large initial random velocities in the P-WDM case
further smooths out the mass distribution, suppressing the formation
of structures with velocity dispersion smaller than the amplitude of
the random velocities.

The structure of the most massive halo in the CDM model, and the
corresponding haloes in WDM and P-WDM, are compared in more detail in
Fig.~\ref{fig:all}. These haloes all have fairly similar total mass,
$M_{200} \sim 1.5 \times 10^{12}$\solarmass\ for both CDM and WDM, and
$M_{200} \sim 1.0 \times 10^{12}$\solarmass\ for P-WDM$_{512}$. The
density profiles of the CDM and WDM runs are nearly indistinguishable
from each other, demonstrating that for this self-consistent WDM model
neither free streaming nor the (relatively small) initial intrinsic
velocities affect the overall structure of the halo. For this choice of WDM
parameters this is expected and has been seen in previous work
\citep[e.g.][]{Bode01,Colin08}.  In contrast, the P-WDM$_{256}$ model,
with its significantly larger intrinsic velocities, does form a core.
Comparison of models P-WDM$_{256}$ and the eight times higher
resolution P-WDM$_{512}$ shows that the
core is numerically well resolved and converged.

The phase space density profiles of models CDM and WDM are also
virtually indistinguishable; they are consistent with a power law, as
found in previous studies (Fig.~\ref{fig:all}, right panel).  Once
again, neither free streaming nor intrinsic velocities affect the
profile. On the other hand, in the P-WDM models a well-resolved core
in the phase space profile develops.  The coarse-grained phase space
density, $F_{\rm NB}$, computed using {\small EnBiD}, has a maximum of
$F_{\rm NB}^{\rm max}\sim 0.5h^2\,M_\odot$~kpc$^{-3}$~(km~s)$^{-3}$,
close to but smaller than the value of $f_{\rm FD}^{\rm
  max}=0.85h^2\,M_\odot$~kpc$^{-3}$~(km~s)$^{-3}$ (shown as the
horizontal red line) calculated from Eqn.(\ref{eq:FD,max}) for a WDM
particle of mass $mc^2=0.03$~keV. This demonstrates that our
simulation methods and our calculation of phase space densities using
{\small EnBiD} are sufficiently accurate to follow correctly the
formation of cores due to the WDM intrinsic velocities. Given the good
numerical convergence of the profiles, we use our three P-WDM$_{256}$
simulations with different initial random phases to obtain statistical
results in more detail below.

In contrast, the fine-grained phase space density bound for the
standard $2$\keV thermal relic is $f^{\rm max}_{\rm FD}(mc^2=2{\rm keV}) =
1.7 \times 10^7$\phaseunit, $\sim 7$ orders of magnitude larger than
for the P-WDM case. The corresponding size of the core is nearly two
orders of magnitude smaller, and is well within the smoothing length
of our simulations. Clearly, as we had anticipated, these simulations
are not able to resolve the core of the WDM model.

\subsection{Physical and phase space density profiles}

\begin{figure*}
\begin{center}
\centerline{\includegraphics[width=1.0\textwidth]{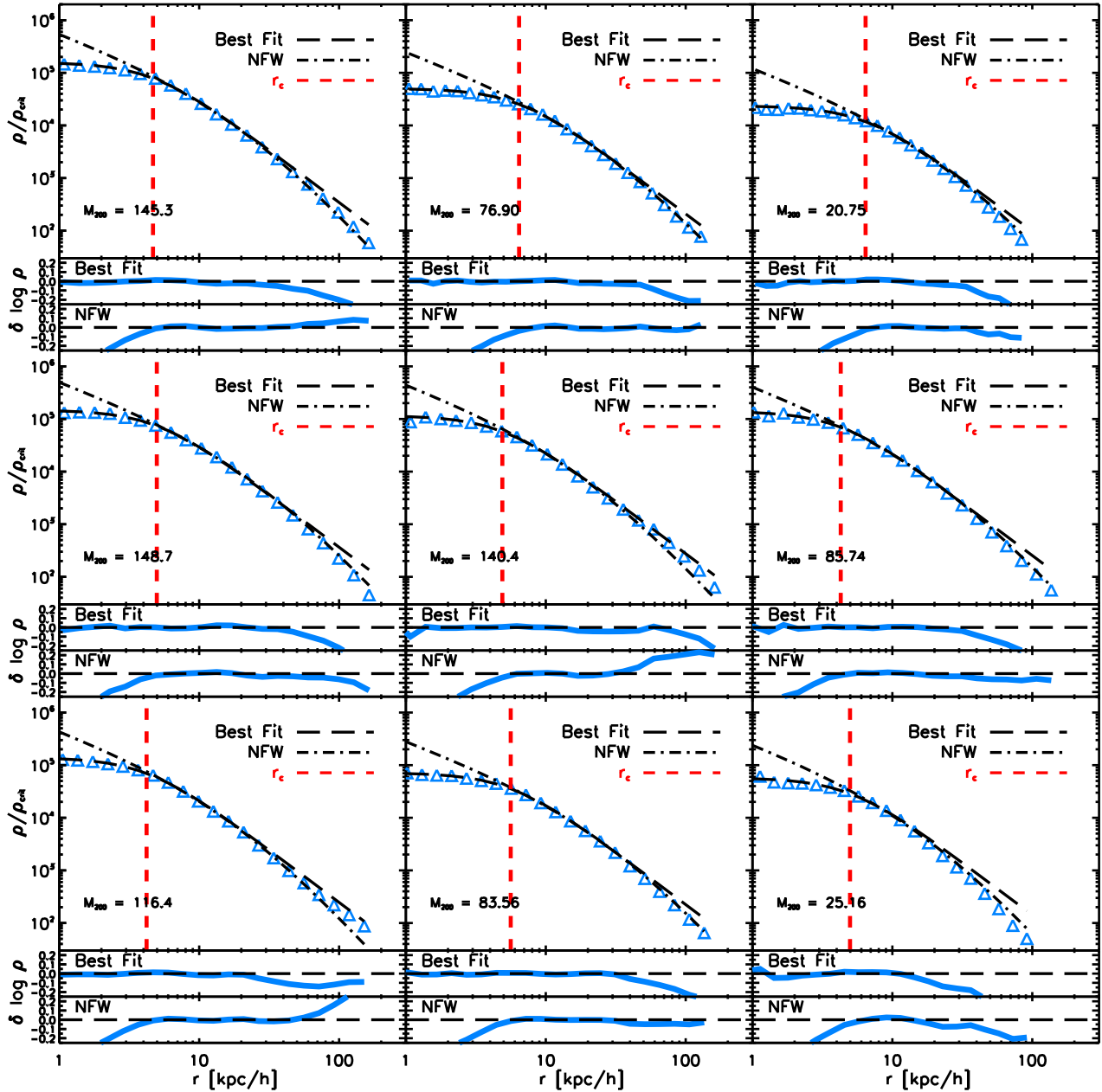}}
\end{center}
\caption{\label{fig:core} Spherically averaged density profiles of
  nine well-resolved haloes in our fiducial P-WDM simulations (blue
  triangles). The top three haloes are chosen from the
  higher-resolution P-WDM$_{512}$ simulation, while the rest are from
  the three P-WDM$_{256}$ simulations. The virial mass, $M_{200}$, in
  units of $\rm 10^{10} h^{-1} M_\odot$, is indicated in each panel.
  Fits using the pseudo-isothermal profiles of Eqn.~(\ref{eq:density})
  over the radial range $2\epsilon < r < r_{200}$ are shown by long
  dashed lines, with the best-fit core radius, $r_c$, indicated by
  vertical red lines. For comparison, the best fit NFW profile over
  the radial range $r_c < r < r_{200}$ is shown as the black
  dot-dashed line. The lower plots show the ratio of the measured
  densities to the best-fit NFW and pseudo-isothermal profiles. All
  nine haloes show a well-resolved core, and their profiles are well
  approximated by the pseudo-isothermal form from well inside $r_c$
  out to $R_{200}$.  }
\end{figure*}

As we have shown in Fig.~\ref{fig:all}, the most massive halo in the
P-WDM simulations develops a central, near uniform density core of
radius a few kiloparsecs. This behaviour is very different from the
cuspy, $\rho(r)\propto 1/r$, profiles familiar from CDM simulations
\citep[hereafter NFW]{Navarro96b,Navarro97}. Fig.~\ref{fig:core} shows
density profiles for a further 9 numerically well-resolved haloes from
the series of P-WDM runs, with masses ranging from $2.1 \times
10^{11}$\solarmass\ to $1.5 \times 10^{12}$\solarmass. All nine haloes
exhibit well-resolved cores and their density profiles are well fit by
the pseudo-isothermal profile for radii well within the core radius
$r_c$ out to $R_{200}$. The isothermal profile fits the halo profiles
near $R_{200}$ to better than 20\%, and even better
further in. Best-fit values for the central core density vary between
the haloes by a factor $\sim 3$.  

\begin{figure*}
\begin{center}
\centerline{\includegraphics[width=1.0\textwidth]{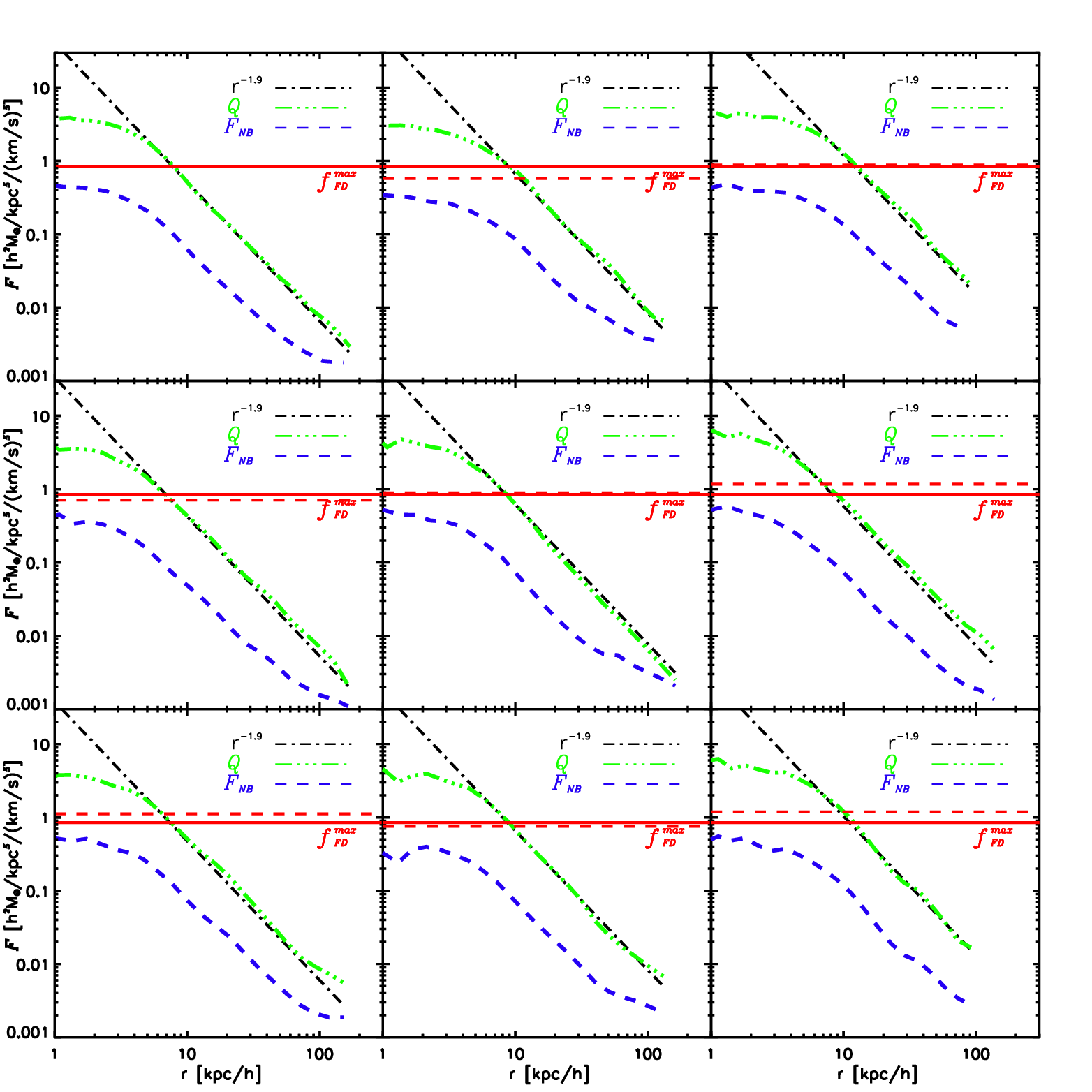}}
\end{center}
\caption{
\label{fig:psdp}
Spherically averaged coarse-grained phase space density profiles for
the nine well-resolved haloes shown in Fig.\ref{fig:core}. The blue
dashed lines show the median values in bins in radius, of the
six-dimensional coarse-grained phase space density profiles, $F_{\rm
  NB}$, calculated using {\small EnBiD}. The pseudo phase space
density estimate, $Q$ (Eqn.\ref{eq:Hogan}), is shown in green, and a
power law, $F\propto r^{-1.9}$ is shown in black. The red solid line
indicates the upper limit to the fine-grained phase space density,
$f^{\rm max}_{\rm FD}$, from Eqn.~(\ref{eq:FD,max}) for
$mc^2=0.03$~keV, the value of $m$ adopted in assigning WDM velocities
in the simulations. The red dashed line indicates the upper limit
$F^{\rm max}_{\rm iso}$ on $F$ calculated from Eqn.~(\ref{eq:iso,max})
using the value of $r_c$ that best fits the density profiles of
Fig.\ref{fig:core} and the velocity dispersion measured for the
particles within $r_c$. These values are very close together. The
maximum value reached by $F_{\rm NB}$ as computed using \small{EnBiD}
(blue dashed line) is close to but smaller than the limits calculated
by either the coarse-grained estimate $F^{\rm max}_{\rm iso}$
calculated from the halo's properties, or the fine-grained phase space
limit, $f^{\rm max}_{\rm FD}$, appropriate for this P-WDM model. This
demonstrates that the simulations satisfy Liouville's theorem, $F_{\rm
  NB}<f^{\rm max}_{\rm FD}$, and that the WDM particle properties can
be estimated from the halo profile, since $F_{\rm NB}^{\rm max}\ltsima
F^{\rm max}_{\rm iso}$. In contrast, the pseudo phase space density,
$Q$, overestimates $F_{\rm NB}$ by a factor of a few for some haloes,
and up to an order of magnitude for others.}
\end{figure*}

Fig.~\ref{fig:psdp} shows the phase space density profiles of the nine
P-WDM haloes represented in Fig.~\ref{fig:core}. The blue
dashed lines show median values in radial bins of the six-dimensional
coarse-grained phase space density profiles, $F_{\rm NB}$, calculated using
{\small EnBiD}. The horizontal red solid lines indicate the
theoretically expected maximum fine-grained phase space density,
$f^{\rm max}_{\rm FD}$, from Eqn.~(\ref{eq:FD,max}) for
$mc^2=0.03$~keV, the WDM mass appropriate to the initial intrinsic
velocities imparted to the particles in the simulations. Clearly, all
the P-WDM haloes have an approximately flat phase space density profile near
the centre. The maximum of this coarse-grained density is very close
to the fine-grained bound appropriate for the $mc^2=0.03$~keV WDM model.

The red horizontal dashed lines in Fig.~\ref{fig:psdp} show $F^{\rm
  max}_{\rm iso}$, the maximum coarse-grained phase space density
estimated from Eqn.~(\ref{eq:iso,max}) assuming a pseudo-isothermal
profile and Maxwellian velocities. To calculate $F^{\rm max}_{\rm
  iso}$ we used the value of $r_c$ which best fits the density
profiles of Fig.\ref{fig:core}, and computed the one-dimensional
velocity dispersion, $\sigma$, for all particles within $r_c$. For all
nine haloes, the coarse-grained estimate, $F^{\rm max}_{\rm iso}$, is
within 30 per cent of $f^{\rm max}_{\rm FD}$. Thus, it is possible to
estimate $f^{\rm max}_{\rm FD}$ by measuring the core radius and
velocity dispersion of the halo.  If real galaxy haloes are dominated
by WDM and baryons have little effect, then our simulations suggests
that it is possible to constrain the mass of a WDM particle, at least
in principle. We apply this idea to dwarf galaxy data in the next
section.

The pseudo phase space density, $Q \equiv \rho/\langle v^2
\rangle^{3/2}$, often used as a proxy for phase space density, is
shown with green lines in Fig.~\ref{fig:psdp}. The shape of $Q(r)$ is
similar to that of $F_{\rm NB}(r)$, but its amplitude is offset to larger
values by up to an order of magnitude. Clearly, using $Q$ to estimate
$F_{\rm NB}$ will result in incorrect constraints on WDM particle masses.

\section{Constraining WDM particle masses from halo profiles}

\begin{figure*}
\begin{center}
\centerline{\includegraphics[width=1.0\textwidth]{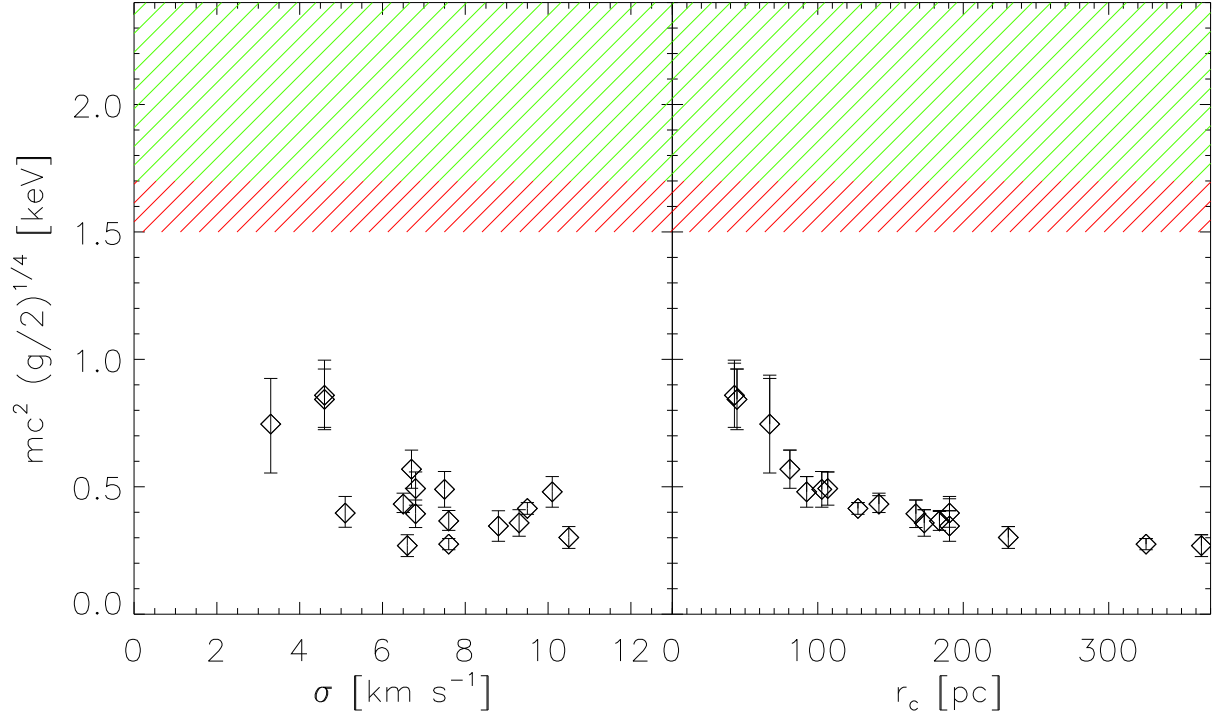}}
\end{center}
\caption{WDM particle mass, $mc^2\,(g/2)^{1/4}$, inferred from the
  satellite properties of Table~2, as a function of the satellite
  velocity dispersion, $\sigma$ ({\em left}), and core radius, $r_c$
  ({\rm right}). For clarity, uncertainties on $\sigma$ and $r_c$ are
  not plotted. The value of $m$ assumes that the coarse-grained phase
  space density, $F^{\rm max}_{\rm iso}$ (Eqn.~\ref{eq:iso,max}),
  equals the theoretical bound, $f^{\rm max}_{\rm FD}$
  (Eqn.~\ref{eq:FD}). This analysis gives values of
  $mc^2(g/2)^{1/4}\approx 0.5$~keV, approximately independently of
  $\sigma$ or $r_c$. Hashed regions indicate the {\em lower} bound on
  $m$ from the Lyman-$\alpha$ forest study of
  \protect\cite{Boyarsky09}, with their 99.7\% and 95\% confidence
  intervals (1.5 and 1.7keV) shown in red and green respectively.}
\label{fig:fig5}
\end{figure*}

\begin{table*}
\begin{tabular}{l|l|l|l|l|l|l}\hline
dSph                    & $R_{h}$     & $\sigma$     & $F_{\rm iso}^{\rm max}$ & $mc^2\,(g/2)^{1/4}$ \\

                                   & pc                        &  km~s$^{-1}$              & $M_{\odot}$~kpc$^{-3}$ (km~s$^{-1}$)$^{-3}$ & keV \\\hline
dSphs from \citet{Gilmore07}\\\hline
Sextans           & $630\pm170$      & $6.6\pm2.3$         & $2.69^{+1.73}_{-1.73}\cdot 10^{-6}$     & $0.269^{+0.043}_{-0.043}$\\\hline
Fornax            & $400\pm103$      & $10.5\pm2.7$        & $4.20^{+2.42}_{-2.42}\cdot 10^{-6}$     & $0.301^{+0.043}_{-0.043}$\\\hline
Leo I             & $330\pm106$      & $8.8\pm2.4$         & $7.36^{+5.13}_{-5.13}\cdot 10^{-6}$     & $0.346^{+0.060}_{-0.060}$\\\hline
Ursa Minor        & $300\pm74$       & $9.3\pm2.8$         & $8.42^{+4.87}_{-4.87}\cdot 10^{-6}$     & $0.358^{+0.052}_{-0.052}$\\\hline
Carina            & $290\pm72$       & $6.8\pm1.6$         & $1.23^{+0.68}_{-0.68}\cdot 10^{-5}$     & $0.394^{+0.054}_{-0.054}$\\\hline
Draco             & $221\pm16$       & $9.5\pm1.6$         & $1.52^{+0.34}_{-0.34}\cdot 10^{-5}$     & $0.415^{+0.023}_{-0.023}$\\\hline
Bootes            & $246\pm28$       & $6.5^{+2.1}_{-1.3}$ & $1.79^{+0.71}_{-0.54}\cdot 10^{-5}$     & $0.432^{+0.043}_{-0.033}$\\\hline
Sculptor          & $160\pm40$       & $10.1\pm0.3$        & $2.73^{+1.37}_{-1.37}\cdot 10^{-5}$     & $0.480^{+0.060}_{-0.060}$\\\hline
Leo II            & $185\pm48$       & $6.8\pm0.7$         & $3.03^{+1.60}_{-1.60}\cdot 10^{-5}$     & $0.493^{+0.065}_{-0.065}$\\\hline
dSphs from \citet{Simon07}\\\hline
Canes Venatici I  & $564\pm36$       & $7.6\pm2.2$         & $2.92^{+0.92}_{-0.92}\cdot 10^{-6}$     & $0.275^{+0.022}_{-0.022}$\\\hline
Ursa Major I      & $318^{+50}_{-39}$& $7.6\pm2.4$         & $9.17^{+4.09}_{-3.67}\cdot 10^{-6}$     & $0.366^{+0.041}_{-0.037}$\\\hline
Hercules          & $330^{+75}_{-52}$& $5.1\pm2.4$         & $1.27^{+0.83}_{-0.72}\cdot 10^{-5}$     & $0.397^{+0.065}_{-0.056}$\\\hline
Leo T             & $178\pm39$       & $7.5\pm2.7$         & $2.97^{+1.68}_{-1.68}\cdot 10^{-5}$     & $0.490^{+0.070}_{-0.070}$\\\hline
Ursa Major II     & $140\pm25$       & $6.7\pm2.6$         & $5.37^{+2.83}_{-2.83}\cdot 10^{-5}$     & $0.569^{+0.075}_{-0.075}$\\\hline
Leo IV            & $116^{+26}_{-34}$& $3.3\pm2.8$         & $1.59^{+1.52}_{-1.64}\cdot 10^{-4}$     & $0.746^{+0.179}_{-0.192}$\\\hline
Coma Berenices I  & $77\pm10$        & $4.6\pm2.3$         & $2.58^{+1.46}_{-1.46}\cdot 10^{-4}$     & $0.843^{+0.119}_{-0.119}$\\\hline
Canes Venatici II & $74^{+14}_{-10}$ & $4.6\pm2.4$         & $2.80^{+1.80}_{-1.64}\cdot 10^{-4}$     & $0.859^{+0.138}_{-0.126}$\\\hline
\end{tabular}
\caption{\label{tab:mass} Parameters for dwarf spheroidal satellites
  of the Milky Way: satellite name (col.~1); 
  half-light radius and velocity dispersion compiled by
  \citet{Boyarsky09} (cols.~2 and~3); coarse-grained limit, $F_{\rm
    iso}^{\rm max}$, derived from the values of $r_h$ and $\sigma$
  (col.~4); mass of the WDM particle assuming that the
  coarse-grained phase space, $F_{\rm iso}^{\rm max}$, equals the
  fine-grained value, $f_ {\rm FD}^{\rm max}$ (col.~5).} 
\end{table*}

We now apply the results of the previous section to the dwarf
spheroidal satellites of the Milky Way. There are numerous claims in
the literature that these galaxies have a central density core
\citep[e.g.][but see \cite{Strigari10}]{Gilmore07,Walker_Penarrubia12}
and there have been attempts to interpret these using phase space
density arguments, often with reference to WDM \citep[e.g.][]{Hogan00,
  Dalcanton01, Bode01, Strigari06, Simon07, Boyanovsky08, Boyarsky09,
  deVega10, deVega12, Maccio12}.  These studies usually assume that the stars in
these galaxies are in dynamical equilibrium and closely trace the dark
matter distribution in their parent dark matter haloes. Constraints on
the mass of the WDM particle are then obtained by comparing the
inferred phase space density with a theoretical expectation, often
based on a proxy such as the pseudo phase space density, $Q$
\cite[e.g.][]{Hogan00}; as we saw before this choice would lead to an
incorrect result.

Our analysis shows that the maximum value of the coarse-grained phase
space density in our N-body simulations is close both to the
fundamental fine-grained bound, $f^{\rm max}_{\rm FD}$
(Eqn.~\ref{eq:FD,max}), and to the estimate, $F^{\rm max}_{\rm iso}$
(Eqn.~\ref{eq:iso,max}), calculated assuming a pseudo-isothermal
profile and Maxwellian velocities. This result allows us to reassess
previous constraints on WDM models and to set new rigorous limits on
the WDM particle mass. In particular, we use the results from our
simulations displayed in Fig.~\ref{fig:psdp} that show that
$F^{\rm max}_{\rm iso}$ differs from $f^{\rm max}_{\rm FD}$ by not
more than about 30\%. We proceed as follows.  

We assume that the observed half-light radius of a dwarf spheroidal
is approximately equal to the radius, $R_h$, at which the projected
dark matter density attains half its central value. The projected
surface density profile, $S(R)$, of the pseudo-isothermal model can be
written as:
\begin{equation} S(R) =
  \int_{-\infty}^{\infty} \rho[(R^2 + z^2)^{1/2}] dz = \frac{S_0
    r_c}{\sqrt{r_c^2 + R^2}}\,,
\end{equation}
where $R$ is projected radius, and $S_0 = \pi \rho_0 r_c$ is the
central surface density. Therefore $R_h=\sqrt{3} r_c$.  We estimate
the central one-dimensional velocity dispersion $\sigma$ of the dark
matter from the (measured) velocity dispersion of the {\em stars}
within $R_h$.  Combining these, we calculate the central maximum phase
space density of the dark matter halo, $F^{\rm max}_{\rm iso}$, using
Eqn.~(\ref{eq:iso,max}).  Finally demanding that $F^{\rm max}_{\rm
  iso}\approx f^{\rm max}_{\rm FD}$, Eqn.~(\ref{eq:Fbound}), yields an
estimate of the WDM particle mass, which we expect to be good to
within 30 per cent. The inferred values are given in the last column
of Table~2 and plotted in Fig.~\ref{fig:fig5}. Of course, these
constraints are only relevant if the satellite dynamics are
collisionless, so that Liouville's theorem is satisfied.

Our constraints on the maximum phase space density and $mc^2g^{1/4}$
differ from other estimates in the literature based on different
methods for estimating the maximum value of the coarse-grained phase
space density. These estimates vary by factors of $\sim50$ to
$\sim0.2$ for $F$, depending on the method used. For example, the
limits from the model of \citet{Hogan00} are a factor of $3.08$ larger
than ours. This disagreement is carried over to the limits set on
$mc^2g^{1/4}$ even when using the same data, by factors ranging from
0.65 to 2.75 (see also the illuminating summary and discussion in
\citet{Boyarsky09}).  In particular, our bounds on $m$ are 1.54 times
larger than those of \citet{Boyarsky09}. Our method has the advantage
over previous methods that it has been explicitly verified by $N$-body
simulations.

The method outlined above yields an estimate of the mass of the WDM
particle, $mc^2(g/2)^{1/4}\approx 0.5$~keV, not a limit on $m$.  This
estimate is based on phase space considerations which, in turn, depend
on the initial intrinsic velocities of the WDM particles. The lower
limit inferred from the Lyman-$\alpha$ forest, $mc^2 \gtsima 1.2$~keV,
on the other hand, is determined by the cutoff in the power spectrum
\citep{Seljak06, Viel08, Boyarsky09}. Although this cutoff is, of
course, induced by free streaming due to the initial velocities, the
two mass estimates exploit different properties of the initial WDM
particle distribution, velocities in one case and the shape of the
power spectrum in the other. For two of the satellites listed in
Table~\ref{tab:mass} -- Coma Berenices~I and Canes Venataci~II -- the
mass estimate from the phase space constraint and the lower limit from
the Lyman-$\alpha$ forest are close. However, for the majority of the
dwarf spheroidals in the table, the WDM particle mass inferred from
the phase space constraint is significantly below the lower limit from
the Lyman-$\alpha$ forest. For these objects the two methods can only
be reconciled if the phase space density is lower than predicted
by Liouville's theorem. This may result, for example, from energy
exchange between the dark matter and the baryons in the halo.

\section{Summary and discussion}
Whether warm dark matter particles decouple as thermal relics or form
from non-equilibrium decay, they acquire initial velocities whose
amplitude depends on the particle's mass. Subsequent free streaming
imposes a cut-off in the primordial spectrum of density perturbations.
In this paper we have performed a series of numerical experiments to
investigate how the intrinsic primordial velocity dispersion of
fermionic dark matter particles affects the central density profile of
the dark matter haloes into which they later collect. For WDM 
the initial velocities are small and resolving them in an
$N$-body simulation of halo formation would require a currently prohibitively
large number of simulation particles. Since we are primarily
interested in the connection between the initial velocities and the
final phase space distribution of the particles, we can circumvent
this problem by decoupling the initial velocities from the free
streaming length. Our simulations therefore do not correspond to a
self-consistent representation of any particular WDM particle
candidate but they are suitable for tackling the problem in hand. The
power spectrum cut-off and primordial velocities we assumed correspond
to those of thermally produced WDM particles of mass of 2~keV and
0.03~keV respectively. For comparison purposes, we also ran
simulations of the standard $\Lambda$CDM case and a self-consistent
WDM model with a particle mass of 2~keV.

Our main results may be summarised as follows: 

\begin{enumerate}

\item Initial particle velocities induce cores in the radial
  profiles of both the physical and the phase space density of dark
  matter haloes. The inner density profile of the simulated haloes is
  well described by the pseudo-isothermal model (with a core).

\item The maximum coarse-grained phase space density of simulated haloes
  \citep[computed using the {\small EnBid} code;][]{Sharma06} is very
  close to the theoretical fine-grained upper bound. This implies that
  it is, in principle,  possible to use phase space arguments to
  constrain the nature of the dark matter. 
\item In contrast, the pseudo phase space density,
  $Q\sim\rho/\sigma^3$, overestimates the maximum phase space density
  by a significant factor, up to an order of magnitude. 

\item Assuming that the velocity distribution of the halo dark matter
  particles is Maxwellian, a simple analytical model that predicts the maximum
  allowed coarse-grained phase space density describes the simulations
  remarkably well (Fig.~3).

\item Application of this analytic model to the kinematical data of dwarf
  spheroidal satellites of the Milky Way, assumed to have a
  pseudo-isothermal density profile with a core, constrains the mass
  of a hypothetical thermal fermionic WDM relic to be
  $mc^2\,(g/2)^{1/4}\approx 0.5$~keV. 
  
\end{enumerate}

The low particle mass we infer from phase space considerations yields
a large free-streaming mass, $M_{\rm FS}\sim 10^{12}\,h^{-1}M_\odot$
(Eqn.\ref{eq:Mfs}). As emphasized by \cite{Maccio12}, this is so
large that a model with this kind of dark matter is unlikely to
produce enough satellites in galaxies like the Milky Way.  Whereas CDM may
suffer from an excess of massive subhaloes -- the \lq too big to
fail\rq\ problem of \cite{Boylan11}, WDM may suffer from a \lq too
small to succeed\rq\ problem.

Our inferred value for $m$ differs from other estimates in the
literature based on different methods to estimate the coarse-grained
phase space density, by factors ranging from 0.2 to 58. Our estimate,
however, is the only one that has been explicitly validated using
cosmological simulations of halo formation. 

The inferred value of $m$ follows from the assumption that the density
profiles of the Milky Way dwarf spheroidal satellites have central
cores. This assumption is controversial: for example
\citet{Walker_Penarrubia12} argue that cores are required by the data,
at least for Fornax and Sculptor, but \citet{Strigari10} have shown
explicitly that the data for these and other dwarfs are consistent
with NFW cusps. In any case, the value of the particle mass required
by the kinematical data, under the assumption that the dwarf
spheroidal haloes do have cores, conflicts with the lower bound on the
WDM particle mass set by observations of the Lyman-$\alpha$ forest
which require the particles to have a mass $m \gtsima 1.2$~keV
\citep{Seljak06, Viel08, Boyarsky09}. This implies that if the cores
are actually real, then they cannot be explained by the free-streaming
velocities of thermally produced WDM particles. Instead, baryonic
processes associated with the forming galaxy, for example, of the kind
originally proposed by \citet{Navarro96a} and more recently seen in
simulations \citep[e.g.][]{Read05, Mashchenko08, Governato10, Brooks12}
would be required.

\section*{Acknowledgements}
It is a pleasure to thank Simon D. M. White, Adrian Jenkins, Alexey
Boyarsky \& Oleg Ruchayskiy, for helpful discussions. We also thank the
referee Andrea Macci{\'o} for a careful reading of the
manscript. Some of simulations used in this work were carried out on
the Lenova Deepcomp7000 supercomputer of the super Computing Centre of
Chinese Academy of Sciences, Beijing, China. LG acknowledges support
from the one-hundred-talents program of the Chinese academy of science
(CAS), the National Basic Research Program of China (program 973 under
grant No. 2009CB24901), {\small NSFC} grants (Nos. 10973018 and
11133003), {\small MPG} partner Group family, and an {\small STFC}
Advanced Fellowship, as well as the hospitality of the Institute for
Computational Cosmology at Durham University. CSF acknowledges a Royal
Society Wolfson Research Merit Award and ERC Advanced Investigator
grant {\small COSMIWAY}. This work was supported in part by an {\small STFC}
rolling grant to the ICC.
\label{lastpage}

\bibliographystyle{mnras}
\bibliography{bibliography}

\end{document}